\documentclass[letterpaper]{article} 
\usepackage{aaai2026}  
\usepackage{times}  
\usepackage{helvet}  
\usepackage{courier}  
\usepackage[hyphens]{url}  
\usepackage{graphicx} 
\usepackage{natbib}  
\usepackage{caption} 
\usepackage[table]{xcolor} 
\usepackage{multirow}  
\usepackage{algorithm}
\usepackage{algorithmic}
\usepackage{booktabs}
\usepackage{makecell}
\usepackage{float}
\usepackage{amsmath}
\usepackage{xcolor}         
\newcommand{\pjc}[1]{\textcolor{black}{#1}}
\usepackage{newfloat}
\usepackage{listings}
\DeclareCaptionStyle{ruled}{labelfont=normalfont,labelsep=colon,strut=off} 
\floatstyle{ruled}
\newfloat{listing}{tb}{lst}{}
\floatname{listing}{Listing}
\title{Deploying Atmospheric and Oceanic AI Models on Chinese Hardware and Framework: Migration Strategies, Performance Optimization and Analysis}
\author{
    Yuze Sun\textsuperscript{\rm 1,2},
    Wentao Luo\textsuperscript{\rm 2},
    Yanfei Xiang\textsuperscript{\rm 1},
    Jiancheng Pan\textsuperscript{\rm 1,3},\\
    Jiahao Li\textsuperscript{\rm 1},
    Quan Zhang\textsuperscript{\rm 1},
    Xiaomeng Huang\textsuperscript{\rm 1}
}
\affiliations{
    \textsuperscript{\rm 1}Department of Earth System Science, Ministry of Education Key Laboratory for Earth System Modelling, Institute for Global Change Studies, Tsinghua University, Beijing, China\\
    \textsuperscript{\rm 2}Huawei Technologies Co., Ltd\\
    \textsuperscript{\rm 3}INSAIT, Sofia University ``St. Kliment Ohridski''

    syz23@mails.tsinghua.edu.cn, hxm@tsinghua.edu.cn

}

\usepackage{bibentry}

\begin{document}

\maketitle

\begin{abstract}
With the growing role of artificial intelligence in climate and weather research, efficient model training and inference are in high demand. Current models like FourCastNet and AI-GOMS depend heavily on GPUs, limiting hardware independence, especially for Chinese domestic hardware and frameworks. \pjc{To address this issue, we present a framework for migrating large-scale atmospheric and oceanic models from PyTorch to MindSpore and optimizing for Chinese chips, and evaluating their performance against GPUs. The framework focuses on software-hardware adaptation, memory optimization, and parallelism. Furthermore, the model's performance is evaluated across multiple metrics, including training speed, inference speed, model accuracy, and energy efficiency, with comparisons against GPU-based implementations. Experimental results demonstrate that the migration and optimization process preserves the models' original accuracy while significantly reducing system dependencies and improving operational efficiency by leveraging Chinese chips as a viable alternative for scientific computing.} This work provides valuable insights and practical guidance for leveraging Chinese domestic chips and frameworks in atmospheric and oceanic AI model development, offering a pathway toward greater technological independence\footnote{https://github.com/zeaccepted/AI-ocean-model-based-on-Mindspore-optimized-for-domestic-chips-in-China}.
\end{abstract}

\section{Introduction}

In recent years, deep learning technology has demonstrated significant potential in atmospheric and oceanic sciences. By learning the spatio-temporal dynamics of climate systems, deep neural networks can efficiently model climate data, enabling high-precision predictions with reduced computational time. In atmospheric science, notable large-scale models include FourCastNet \citep{pathak2022fourcastnet}, GraphCast \citep{lam2023learning}, Fuxi \citep{chen2023fuxi}, Fengwu \citep{chen2023fengwu}, and Pangu-Weather \citep{bi2023accurate}. In oceanography, key models include AI-GOMS \citep{xiong2023ai}, Xihe \citep{wang2024xihe} Langya \citep{yang2024langya} and Wenhai \citep{cui2025forecasting}. These models leverage atmospheric and oceanic reanalysis data, utilizing high-dimensional feature extraction techniques to generate accurate predictions rapidly. However, efficiently training and running these deep learning models demands substantial computational resources. Currently, \pjc{Graphics Processing Units (GPUs)} dominate as the primary hardware platform due to their powerful parallel computing capabilities, which are essential for deep learning. Yet, GPU-based solutions face challenges, including high costs, supply chain dependencies, and energy inefficiency for certain tasks.

Against this backdrop, Chinese chips and frameworks are gaining prominence. Chinese artificial intelligence (AI) acceleration chips, such as Sugon’s \pjc{Deep Computing Unit (DCU)} and Huawei’s Ascend, are becoming key platforms for deep learning training and inference due to their high cost-effectiveness and maturing software ecosystems. Using Chinese chips for model training and inference in meteorological and oceanic AI can reduce reliance on foreign hardware and advance high-performance computing and autonomous control technologies. Additionally, deep learning frameworks, core tools for large-scale model innovation, play a crucial role in development. Leading Chinese frameworks such as Huawei’s MindSpore \citep{chen2021deep} and Baidu’s PaddlePaddle \citep{ma2019paddlepaddle} can maximize computational potential and enhance efficiency when optimized for synergy with Chinese chips.

Although Chinese chips and frameworks have made significant progress in AI computing, the current software and hardware ecosystem is still primarily dominated by mainstream GPUs and their supported frameworks (e.g., \pjc{PyTorch\footnote{https://pytorch.org/}, TensorFlow\footnote{https://www.tensorflow.org/}}). The Chinese deep learning ecosystem faces several key challenges compared to traditional frameworks. First, atmospheric and oceanic models are typically developed using frameworks like PyTorch, which are not fully compatible with MindSpore\footnote{https://www.mindspore.cn/} regarding operators. Direct migration may lead to missing operators or reduced computational efficiency, challenging migration. Second, the hardware architecture of Chinese chips significantly differs from GPU-based ones, necessitating the redesign of many optimization strategies to suit chip-specific features. Finally, the ecosystem of Chinese chips is still immature, lacking a rich library of functions and debugging tools, which adds extra challenges to development and optimization. These issues make it difficult to directly run atmospheric and oceanic models on Chinese chips. Therefore, \textit{how to migrate from PyTorch to Chinese frameworks, optimize operator adaptation and hardware utilization, and ensure efficient model operation on Chinese chips has become a critical research topic.}

Some studies have already analyzed the performance and optimization of MindSpore on Ascend. \cite{lu2022evaluation} conducted a detailed comparison of PyTorch, TensorFlow, and MindSpore on Ascend Neural Processing Unit (NPU) and GPU. \cite{wang2023analysis} performed extensive experiments on Ascend NPUs using MindSpore, analyzing key features related to performance and optimization based on experimental results. \cite{zhu2023performance} conducted comprehensive evaluations at both the model and operator levels on Ascend NPUs, comparing the performance of MindSpore and PyTorch. In addition, some studies \citep{deng2024implementation,mu2024devmut,yin2024image,yang2025research}, also focus on the migration performance of framework chips in the fields of computer vision or large language models. However, most of these studies use standard deep learning networks and data sets for testing, such as \pjc{Convolutional Neural Network (CNN) and Transformer, etc.}, and there is still no research on migration assessment tests based on large atmospheric and ocean deep learning models.

This study addresses existing challenges by developing a model migration and optimization framework that transfers large-scale atmospheric and oceanic PyTorch-based models to the MindSpore framework. It then systematically evaluates Chinese hardware platforms (DCU and Ascend).
\pjc{In migration and optimization framework, we focuses on:} (1) Model Migration—adapting PyTorch code to MindSpore while ensuring functional correctness;  (2) Performance Optimization—leveraging Chinese chips' hardware features to refine operators and implement distributed training strategies;  and (3) Performance Evaluation—comparing Chinese chips with GPUs across metrics like training/inference speed, accuracy, and energy efficiency to assess their suitability for climate modeling tasks. By enabling large-scale model deployment on Chinese platforms, this work advances the localization of atmospheric and oceanic simulations while validating DCU and Ascend's capabilities in high-performance scientific computing.  The findings expand Chinese chips' application scope, offer technical guidance for cross-framework model migration, and provide critical data to support the development of Chinese deep learning hardware and software ecosystems. \pjc{We summarize the main contributions:
\begin{itemize}
\item We developed a migration and optimization framework that adapts large-scale atmospheric and oceanic models from PyTorch to MindSpore with functional correctness;
\item We performed systematic optimization and evaluation on Chinese chips (DCU and Ascend), comparing them with GPUs in terms of speed, accuracy, and energy efficiency;
\item We advanced the localization of scientific computing, extended the use of Chinese hardware, and provided technical guidance and empirical evidence to support the Chinese AI ecosystem.
\end{itemize}}

\section{Method}
\begin{figure*}[t]
\centering
\includegraphics[width=0.85\linewidth]{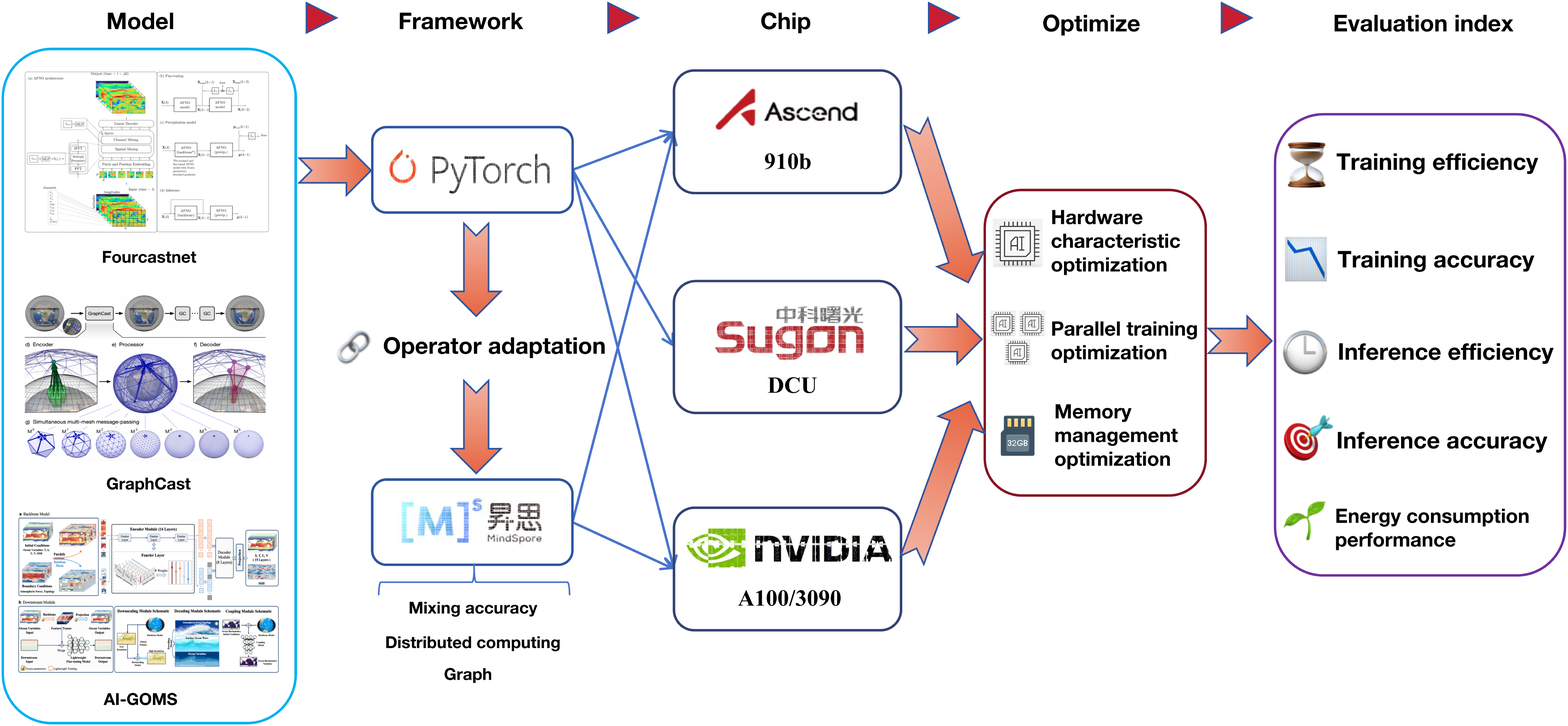} 
\caption{The migration and optimization framework for atmospheric and oceanic models from PyTorch to MindSpore based on different hardware.}
\label{fig1}
\end{figure*}

Figure \ref{fig1} systematically presents the complete process of chip framework migration assessment. The evaluation system includes the migration of large atmospheric and oceanic models to the Mindspore framework, hardware chip adaptation and optimization strategies, and multi-dimensional evaluation index design.

\subsubsection{Model Migration Framework Design.}
Due to significant differences in framework design and operator support, direct migration often encounters issues such as model structure incompatibility, missing operators, and insufficient performance optimization. To address these challenges, we designed a step-by-step migration workflow (Figure S2), including key processes such as model structure adjustments, operator adaptation and replacement, and optimizations leveraging MindSpore-specific features, with AI-GOMS as the primary example.

Regarding model structure adjustments, PyTorch employs a dynamic graph mechanism, allowing model structures to be modified dynamically during runtime, whereas MindSpore adheres to a static graph mechanism. This requires a complete redesign of the model's overall logic during migration. For AI-GOMS, the model architecture had to be redefined in MindSpore using a static computation graph to ensure compatibility and trainability. Additionally, MindSpore's optimization for static graphs requires developers to declare input dimensions and shapes explicitly. As a result, input tensors needed to be pre-defined during model initialization.
Regarding operator adaptation, MindSpore's operator library does not yet comprehensively cover commonly used PyTorch operators, making operator incompatibility a common issue during migration. For example, AI-GOMS relies on Fourier convolution functions like `torch.nn.fft' in PyTorch.

To address this, we adopted the following strategies: (1) prioritizing the use of equivalent operators already available in MindSpore as replacements for PyTorch operators; (2) developing custom operators to extend MindSpore's operator library for unsupported operations; and (3) restructuring computational logic or utilizing low-level APIs to implement the functionality of certain complex operators. Regarding optimizations leveraging MindSpore-specific features, we entirely use unique functionalities provided by the MindSpore framework, such as mixed precision training, built-in distributed computation support, and graph mode optimizations. Mixed precision training enables significant reductions in memory usage while maintaining model accuracy, thereby improving computational efficiency. Distributed computation support allowed us to split large models across multiple Ascend chips for training, effectively addressing the limitations of single-chip computing power. Graph mode optimization analyzes the static computation graph during the compilation phase, improving training speed by optimizing the computational graph.

\subsubsection{Optimization Strategies for Adapting to Chips.}
After migrating the model to MindSpore, we designed a series of targeted optimization strategies to fully exploit the hardware capabilities of Chinese chips (DCU and Ascend). These optimizations include hardware-specific enhancements, distributed training optimizations, and memory management improvements.

The Ascend 910b, as a high-performance AI chip, offers computational power that rivals or surpasses high-end GPUs in specific scenarios. To capitalize on this, we utilize Ascend's built-in hardware accelerators (such as matrix computation units and AI operator optimization engines) to optimize the model at the operator level. For instance, using Ascend's matrix acceleration engine significantly improves computational efficiency in the matrix multiplication operations of AI-GOMS. Additionally, we leveraged MindSpore's built-in operator library for convolution and normalization operations to perform hardware instruction-level optimizations, reducing data transfer and computational latency. During the training of large models, the substantial number of parameters and high computational complexity often make it challenging for a single chip to handle training efficiently. To address this, we utilized MindSpore's distributed training support to divide the training data into parts processed by multiple Ascend chips in parallel under a model parallelism mode. This implementation of distributed training significantly reduced single-chip training time while enhancing overall training efficiency. In memory management, the memory usage of deep learning models during training and inference is a critical factor affecting hardware performance. We implemented specific memory allocation and computation scheduling strategies to optimize memory usage on DCU and Ascend. On the one hand, we leveraged MindSpore's mixed precision training functionality to reduce some model weights from 32-bit floating-point (FP32) to 16-bit floating-point (FP16), significantly lowering memory overhead. On the other hand, we utilized Ascend's pipelined execution features to schedule the forward and backward propagation stages in a phased manner, avoiding excessive memory usage peaks.

\subsubsection{Evaluation Benchmarks.} 
Several key metrics are defined to evaluate model performance after migration. Training efficiency is assessed by recording per-epoch and total training time. Inference efficiency is measured through single-chip inference time, with cross-platform comparisons. Following \citep{rasp2020weatherbench}, accuracy is evaluated using RMSE and ACC to analyze atmospheric and oceanic variable predictions over time. Energy performance is monitored during training and inference using hardware tools, with energy efficiency calculated \citep{bridges2016understanding} for comprehensive platform comparisons. The corresponding formulas are shown in the appendix.

\section{Experimental Setup}

\subsubsection{Hardware Configuration.}
The experimental hardware platform includes Chinese chips and mainstream GPUs for performance comparison in deep learning tasks. The table \ref{tab1} details the configurations. The Huawei Ascend 910b chip delivers 256 TFLOPS (FP16), 32GB high bandwidth memory, and a 1.25 GHz clock speed, supporting MindSpore and PyTorch frameworks. Both single-card and multi-card setups were tested for distributed training and inference. The Sugon DCU Z100L chip, based on the ROCm architecture, emphasizes low power consumption and high performance. It requires precompiled PyTorch packages from Sugon’s platform, as official PyTorch is incompatible. As the DCU platform currently does not support the MindSpore framework, no experiments based on MindSpore framework are conducted on this chip. Experiments cover its 16GB and 32GB memory variants. The GPU platform includes NVIDIA A100 and NVIDIA 3090 GPUs, with the runtime environment based on the CUDA architecture. The GPU platform serves as a control group for comparison of training time, inference speed, and accuracy performance. All hardware platforms were configured with the same version of the operating system and software environment to ensure fair comparisons between platforms. 

\begin{table}[h]

\centering
\resizebox{0.98\linewidth}{!}{
\begin{tabular}{ccccc}
\toprule
\textbf{Hardware} & 
\makecell{\textbf{Memory} \\ \textbf{(GB)}} & 
\makecell{\textbf{FP16} \\ \textbf{(TFlops)}} & 
\makecell{\textbf{FP32} \\ \textbf{(TFlops)}} & 
\makecell{\textbf{Bandwidth} \\ \textbf{(GB/s)}} \\
\midrule
Ascend 910b & 40GB & 376 & 94 & 392 \\
DCU Z100L   & 32GB & 24.5 & 12.2 & 1024 \\
Nvidia A100 & 40GB & 312 & 19.5 & 1597 \\
Nvidia 3090 & 24GB & 35.58 & 35.58 & 936 \\
\bottomrule
\end{tabular}}
\caption{Hardware Configuration}\label{tab1}
\end{table}

\subsubsection{Datasets and Models.}

We use the ERA5 \citep{hersbach2020era5} and HYCOM \citep{chassignet2007hycom} global reanalysis datasets. ERA5 provides 0.25° resolution atmospheric variables (temperature, humidity, wind speed) and oceanic surface data, while HYCOM offers oceanic parameters (salinity, temperature, currents) across depth layers, interpolated from 0.08° to 0.25° resolution. \pjc{To ensure comparability across platforms, the data is first standardized through normalization and interpolation, then split into training, validation, and testing subsets.} This experiment selects three representative models from the atmospheric and oceanic domains, FourCastNet, GraphCast, and AI-GOMS, encompassing diverse deep learning architectures such as Transformer, Graph Neural Network (GNN), and  Adaptive Fourier Neural Operator (AFNO), as well as a variety of task scenarios.

FourCastNet \citep{pathak2022fourcastnet} is a deep learning weather forecasting model. Based on AFNO \citep{guibas2021adaptive}, it achieves efficient multivariate meteorological predictions through training on large-scale reanalysis datasets. FourCastNet offers superior parallel processing capabilities and computational efficiency, significantly accelerating inference speed compared to traditional numerical weather prediction methods. However, its training process remains computationally intensive - the original study reports requiring 16 hours on 64 NVIDIA A100 GPUs for end-to-end training. This model is an effective benchmark for evaluating Chinese chips' performance handling atmospheric data operations, particularly Fourier transforms.

GraphCast \citep{lam2023learning} is a GNN-based weather forecasting model \citep{alet2019graph} that effectively captures meteorological data's non-Euclidean characteristics, including flow fields and dynamic spatial relationships. Through sophisticated node-edge representations, it models large-scale atmospheric interactions with superior performance. However, this comes with significant computational demands - training requires about four weeks on 32 TPUs. We use GraphCast to evaluate the chips' capability in handling non-Euclidean data structures and computations.

AI-GOMS \citep{xiong2023ai} is the world's first large-scale oceanic model. It combines Fourier-encoded autoregressive pretraining \citep{he2022masked,guibas2021adaptive} with lightweight fine-tuning to predict key global oceanic variables at 0.25° spatial and 1-day temporal resolution across 15 depth layers for 30-day forecasts. The model significantly improves efficiency and accuracy for operational ocean forecasting and research. Training requires approximately 34.2 hours on 16 NVIDIA A100 GPUs. We use AI-GOMS to evaluate Chinese chips' performance in processing Fourier-encoded oceanic data.


\subsubsection{Training and Inference Configuration.}
We evaluate model performance across frameworks and hardware platforms by measuring: (1) time efficiency, (2) training accuracy, and (3) energy consumption. To ensure fair comparisons:
\begin{itemize}
\item All experiments use identical hyperparameters (learning rate, optimizer, batch size)
\item A fixed random seed guarantees reproducibility
\item Training duration is determined by convergence (when loss stabilizes)
\item Mixed precision training is enabled on GPU/Ascend platforms, with accuracy validated against full precision
\item Inference tests use pre-trained weights on single cards to assess speed and accuracy
\end{itemize}

\begin{table*}[h]
\begin{center}
\resizebox{0.9\linewidth}{!}{
\begin{tabular}{ccccccc}
\toprule
$Chip$&  $Memory$ & $Number$ & $batch\_size$&  $time/epoch(s)$& $Convergence$ $epoch$ $number$&$Training$ $time(s)$\\
\midrule
         A100&  40G&16&2&  5400&  75&405000\\ 
         A100&    40G&16&1&  8800&  60&528000\\ 
         3090&    16G&16&1&  10200&  65&663000\\ 
         \rowcolor{blue!10}
         910b&    40G&16&2&  5580&  75&418500\\ 
         \rowcolor{blue!10}
         910b&    40G&16&1&  8960&  60&537600\\ 
         \rowcolor{blue!10}
         DCU&  32G&16& 1& 12600& 60&756000\\ 
         \rowcolor{blue!10}
         DCU&  32G&32& 1& 6700& 75&502500\\
         \rowcolor{blue!10}
         DCU&  32G&64& 1& 4500& 75&337500 \\
\bottomrule
\end{tabular}}
\end{center}
\caption{Comparison of AI-GOMS training efficiency with PyTorch framework}\label{tab:t2}
\end{table*}

\begin{table*}[h]

\begin{center}
\resizebox{0.9\linewidth}{!}{
\begin{tabular}{ccccccc}
\toprule
$Chip$&  $Memory$ & $Number$ & $batch\_size$&  $time/epoch(s)$& $Convergence$ $epoch$ $number$&$Training$ $time(s)$\\
\midrule
         A100&  40G&16&2&  5500&  75&412500\\ 
         A100&    40G&16&1&  8960&  60&537600\\ 
         \rowcolor{blue!10}
         910b&    40G&16&2&  5400&  70&378000\\ 
         \rowcolor{blue!10}
         910b&    40G&16&1&  8800&  55&484000 \\
\bottomrule
\end{tabular}}
\end{center}
\caption{Comparison of AI-GOMS training efficiency with MindSpore framework}\label{tab:t3}
\end{table*}

\section{Experimental Results}

\subsubsection{Training Efficiency Analysis.}
Training time is a key indicator for assessing hardware platform performance. We conducted complete training of FourCastNet, GraphCast, and AI-GOMS on different hardware platforms and frameworks, training the models until full convergence. Each model's average training time per epoch and total training duration were recorded. By comparing the performance of various platforms, we evaluate Chinese chips' training efficiency and adaptability.

Using the AI-GOMS model as an example, Table \ref{tab:t2} shows that in FP16 precision mode, the Ascend 910b platform running on the PyTorch framework achieves training times nearly identical to the A100 platform, with a difference of less than 5\% in training time per epoch. This result demonstrates that Ascend 910b offers hardware performance comparable to mainstream GPUs. The training time for AI-GOMS on the DCU platform is relatively longer, with the total training time reaching approximately 1.29 times that of the A100 platform. This performance difference likely arises from constraints in computing power and memory bandwidth. However, the training time significantly decreases with multi-device parallel training, highlighting its scalability potential. After migrating to the MindSpore framework, Ascend 910b exhibits higher efficiency when leveraging its distributed training features, as shown in Table \ref{tab:t3}. With the optimization features provided by MindSpore, the total training time on the Ascend 910b platform drops by approximately 10\% compared to the PyTorch framework. Additionally, the distributed training mode of MindSpore enhances resource utilization and fully exploits the hardware acceleration advantages of the Ascend platform.

\subsubsection{Training Accuracy Analysis.}
During the deep learning model training phase, training accuracy is typically evaluated by monitoring the loss function and the decline in convergence speed. For large-scale models like AI-GOMS, which involve massive datasets and high parameter counts, the convergence speed and the final loss value are critical indicators of model performance after migration.

Using the AI-GOMS model on the PyTorch framework as an example, the training conducted on the Ascend 910b and DCU platforms shows a decline in loss function consistent with GPU platforms such as A100. The experiments reveal that the loss function decreases rapidly during the first 20 epochs and gradually stabilizes after 40 epochs. Across different hardware platforms, the overall shape of the loss curves remains similar, indicating that the migrated model maintains consistent training performance on multiple platforms. The training process demonstrates higher stability on the MindSpore framework running on the Ascend 910b platform. The loss curve is smoother than the PyTorch framework, without noticeable fluctuations or instability. This indicates that the migration to the MindSpore framework achieves good compatibility with the optimizer and operators, effectively leveraging the hardware features of the Ascend platform. Further analysis shows that the MindSpore framework, through static graph optimization and mixed precision training, efficiently handles parameter updates and gradient computations for large-scale models, thereby improving training stability and efficiency. The visualization of the training loss is shown in Figure S1.

\subsubsection{Inference Efficiency Analysis.}
We evaluate the time consumption for single inference tasks across different models, frameworks, and hardware platforms in the inference time tests. The results are shown in Table \ref{tab:t4}. Using AI-GOMS as an example, the single-step inference time on the Ascend 910b platform is comparable to that on the A100 platform, while the inference time on the DCU platform is slightly higher than that on the A100. Notably, when the model is migrated to the MindSpore framework, the single-step inference speed on the Ascend 910b platform improves compared to the PyTorch version, indicating that MindSpore offers particular advantages in optimizing for the Ascend chips. For other models, the inference time for FourCastNet and GraphCast shows more significant differences across hardware platforms. For instance, the inference time of FourCastNet on the Ascend 910b platform is slightly higher than that on the A100 but still outperforms the DCU. In contrast, GraphCast, due to the computational complexity of its graph neural network structure, exhibits higher inference times on both the Ascend 910b and DCU platforms compared to the A100. These results suggest that while the Ascend and DCU platforms demonstrate competitiveness in certain tasks, further optimization is required for specific types of deep learning models, such as graph neural networks.

\subsubsection{Inference Accuracy Analysis.}
We compare the inference results obtained using the MindSpore framework on the Ascend 910b platform with those obtained using the PyTorch framework on the A100 platform, focusing on the output accuracy of several key variables (e.g., $U_0$, $T_0$, $S_0$). The results indicate that most variables maintain consistent inference accuracy across the two frameworks, though some variables exhibit slight differences in accuracy.

\begin{figure}[t]
 \centerline{\includegraphics[width=0.95\linewidth]{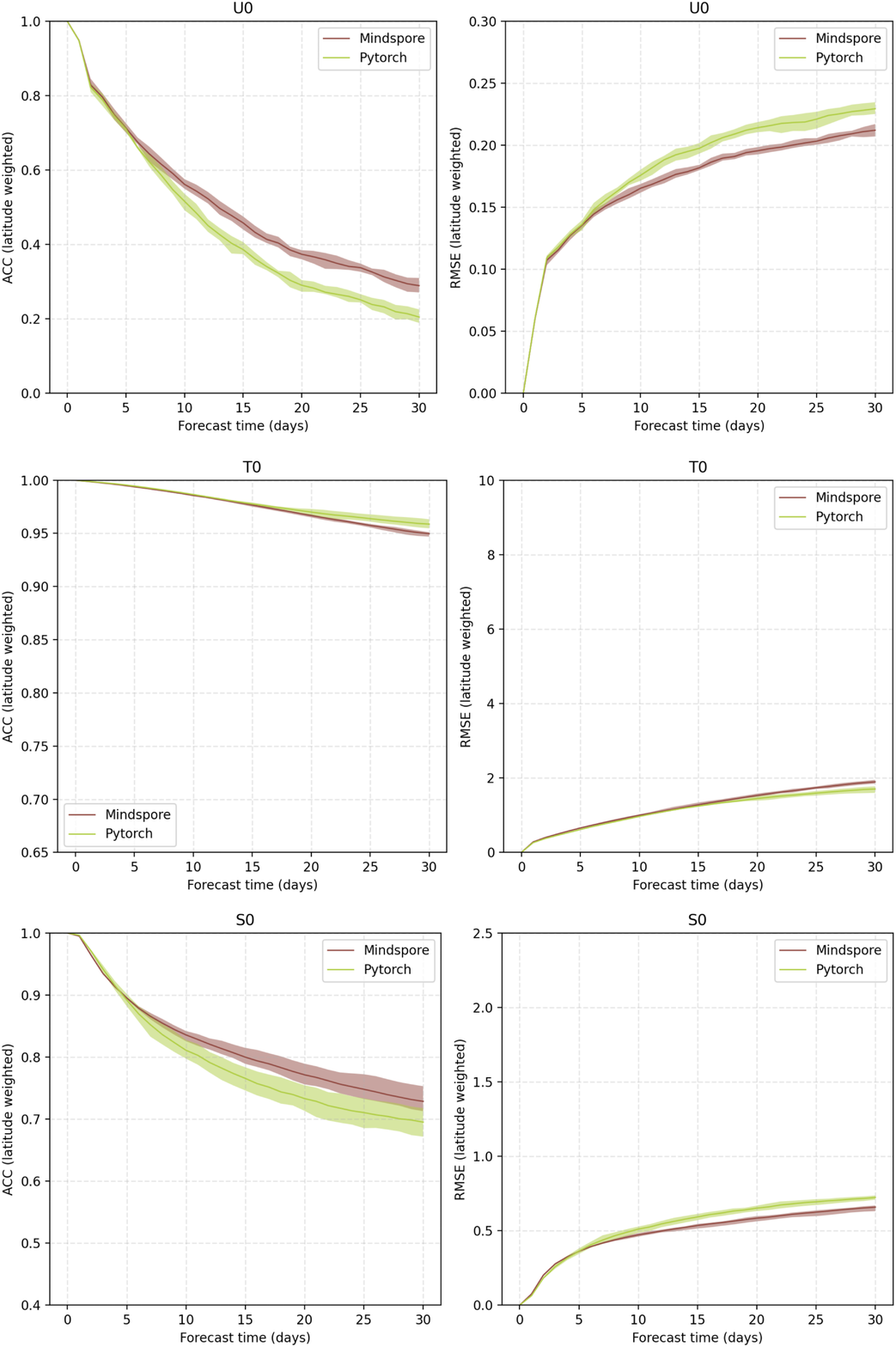}}
  \caption{Comparison of Inference Accuracy between MindSpore and PyTorch Frameworks}\label{fig3}
\end{figure}

As shown in Figure \ref{fig3}, the predictions for $U_0$ (horizontal velocity component) demonstrate strong consistency between the MindSpore and PyTorch frameworks, with high inference accuracy, indicating that the model's dynamic characteristics are well-preserved after migration. Similarly, the predictions for $S_0$ show high similarity, with minimal errors between the two frameworks. However, for $T_0$, the predictions from the MindSpore framework exhibit a certain degree of error compared to those from the PyTorch framework, reflecting slightly lower inference accuracy. This discrepancy might be attributed to the numerical characteristics of the temperature field variable, such as its sensitivity to spatial distribution. Additionally, the error could result from numerical approximations in operator implementations or differences in handling floating-point computations.


\begin{table}[t]
\begin{center}
\resizebox{0.9\linewidth}{!}{
\begin{tabular}{ccccc}
\toprule
$Model$ & $Framework$ & $Chip$ & $test$ $time/step(s)$ \\
\midrule
\multirow{3}{*}{Fourcastnet} & Pytorch & A100 & 0.353 \\
                             & Pytorch & 910b & 0.389 \\
                             & Pytorch & DCU  & 0.400 \\
\midrule
\multirow{2}{*}{Graphcast} & Pytorch & A100 & 1.035 \\
                           & Pytorch & 910b & 1.169 \\
\midrule
\multirow{5}{*}{AI-GOMS} & Pytorch   & A100 & 0.675 \\
                         & Pytorch   & 910b & 0.680 \\
                         & Pytorch   & DCU  & 0.705 \\
                         & Mindspore & A100 & 0.598 \\
                         & Mindspore & 910b & 0.605 \\
\bottomrule
\end{tabular}}
\end{center}
\caption{Comparison of inference efficiency of different models on different chips and frameworks}
\label{tab:t4}
\end{table}

\subsubsection{Energy Consumption Analysis.}
Energy consumption is a critical performance metric for Chinese chips, particularly in scenarios with high energy-saving requirements. In this experiment, hardware monitoring tools were used to record the power consumption of each platform during training and inference tasks, and the relative energy efficiency ratio was calculated, as shown in Table S1. Using AI-GOMS as an example, during the training phase, the Ascend 910b platform demonstrated superior energy efficiency compared to GPUs, with an average power consumption increase of approximately 15\% while maintaining comparable performance to GPUs, resulting in a significant improvement in energy efficiency ratio. The DCU platform exhibited even lower energy consumption, but due to its relatively weaker computational power, its overall energy efficiency ratio was comparable to GPUs. In the inference phase, the DCU platform achieved the lowest power consumption, with an energy efficiency ratio 1.3 times higher than GPUs, making it particularly suitable for long-duration inference tasks. The Ascend 910b platform consumed 10\% less power than GPUs during inference tasks while maintaining high performance levels. Overall, Chinese chips exhibited outstanding energy efficiency, particularly in inference scenarios. The DCU platform showed a significant advantage in energy savings, while the Ascend 910b platform demonstrated a strong ability to balance energy consumption and performance.

\section{Conclusion and Discussion}

This study demonstrates the feasibility and advantages of migrating large-scale atmospheric and oceanic models from PyTorch to MindSpore for efficient execution on Chinese hardware platforms such as Ascend 910b and DCU. Through systematic experimentation, we confirm that the migrated models retain comparable accuracy to their original PyTorch counterparts, with deviations under 5\%, while achieving competitive computational performance. Notably, Ascend 910b exhibits strong capabilities in distributed training scenarios, delivering GPU-level throughput with 10–15\% lower energy consumption. In contrast, DCU shows particular promise in energy-efficient, real-time inference tasks due to its superior power efficiency.

However, challenges remain in optimizing operator support for complex model architectures and further bridging the accuracy gap in high-resolution, computationally intensive simulations. Future work should focus on enhancing MindSpore’s operator library, refining distributed training efficiency, and exploring hardware-software co-design strategies tailored to specific meteorological and oceanographic applications. Additionally, fostering a robust ecosystem—including optimized toolchains, open-source collaboration, and industry integration—will be critical to advancing the adoption of Chinese platforms in large-scale scientific deep learning.

Ultimately, this research underscores the significant potential of Chinese chips in climate and ocean modeling, paving the way for further innovations in performance, energy efficiency, and scalability. Continued advancements in framework optimization, hardware specialization, and ecosystem development will be essential to fully realize this potential and strengthen the competitiveness of Chinese solutions in high-performance AI-driven scientific computing.

\onecolumn
\newpage
\twocolumn
\bibliography{aaai2026}
\newpage

\onecolumn
\section{Appendix}

\subsection{Formula}
The RMSE for the variable c at forecast time step t is defined as
\begin{equation}
\text{RMSE}(c, t) = \sqrt{\frac{1}{|G|} \sum_{i,j} L(i) \left( X_{\text{pred}}^{t}[c, i, j] - X_{\text{true}}^{t}[c, i, j] \right)^2},
  \end{equation}

where \( X^{t}_{\text{pred}} \) and \( X^{t}_{\text{true}} \) represent the value of predicted and true variable \( c \) at coordinates \( (i, j) \). \(L(i)\) is the latitude-dependent weighting factor at coordinate \( (i, j) \).

The ACC for the variable c at forecast time step t is defined as
\begin{equation}
\text{ACC}(c, t) = \frac{\sum_{i,j} L(i) \tilde{X}^{t}_{\text{pred}}[c, i, j] \tilde{X}^{t}_{\text{true}}[c, i, j]}{\sqrt{\sum_{i,j} L(i) \left( \tilde{X}^{t}_{\text{pred}}[c, i, j] \right)^2 \sum_{i,j} L(i) \left( \tilde{X}^{t}_{\text{true}}[c, i, j] \right)^2}}.
  \end{equation}

where \( \tilde{X}^{t}_{\text{pred}} \) represents the climatological-mean-subtracted value of the predicted and true variable \( c \) at coordinates \( (i, j) \).

The energy efficiency ratio is calculated as
\begin{equation}
\text{Energy Efficiency Ratio} = \frac{\text{Throughput}}{\text{Power Consumption}} = \frac{T(c,t)}{P(c,t)}
  \end{equation}
where \( T(c,t) \) represents the throughput at coordinates \( (c,t) \), and \( P(c,t) \) represents the power consumption at coordinates \( (c,t) \) .

\renewcommand{\thefigure}{S\arabic{figure}} 
\renewcommand{\thetable}{S\arabic{table}} 
\setcounter{figure}{0} 
\setcounter{table}{0}  

\subsection{Table}

\begin{table*}[h]
\centering
\resizebox{0.95\linewidth}{!}{
\begin{tabular}{ccccc}
\toprule
$Chip$& \makecell{$Average$ $power$ $consumption$ \\$in$ $training$ $stage(W)$} & \makecell{$Energy$ $efficiency$ $ratio$ \\$in$ $training$ $stage(\%)$} & \makecell{$Average$ $power$ $consumption$ \\$in$ $inference$ $stage(W)$}&\makecell{$Energy$ $efficiency$ $ratio$\\ $in$ $inference$ $stage(\%)$}\\
\midrule
    A100 & 235 & 100 & 100 & 100 \\
    910b & 200 & 115 & 90 & 110 \\
    DCU  & 180 & 91 & 70 & 137\\
\bottomrule
\end{tabular}}
\caption{Energy Consumption Analysis}

\end{table*}

\subsection{Figure}
\begin{figure*}[h]
 \centerline{\includegraphics[width=26pc]{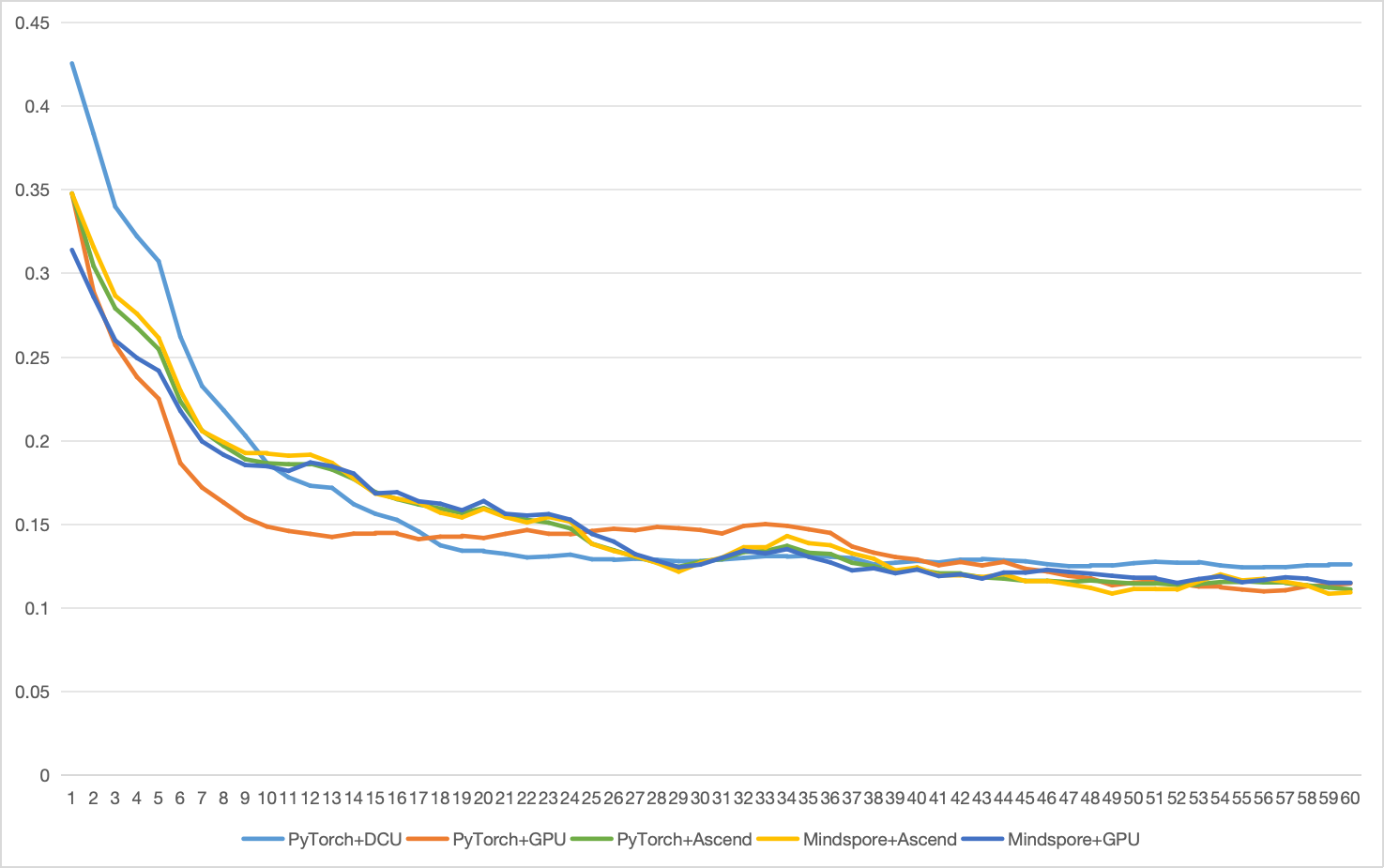}}
  \caption{Training Performance of AI-GOMS under different chip and frameworks}
\end{figure*}

\begin{figure*}[h]
 \centerline{\includegraphics[width=0.95\linewidth]{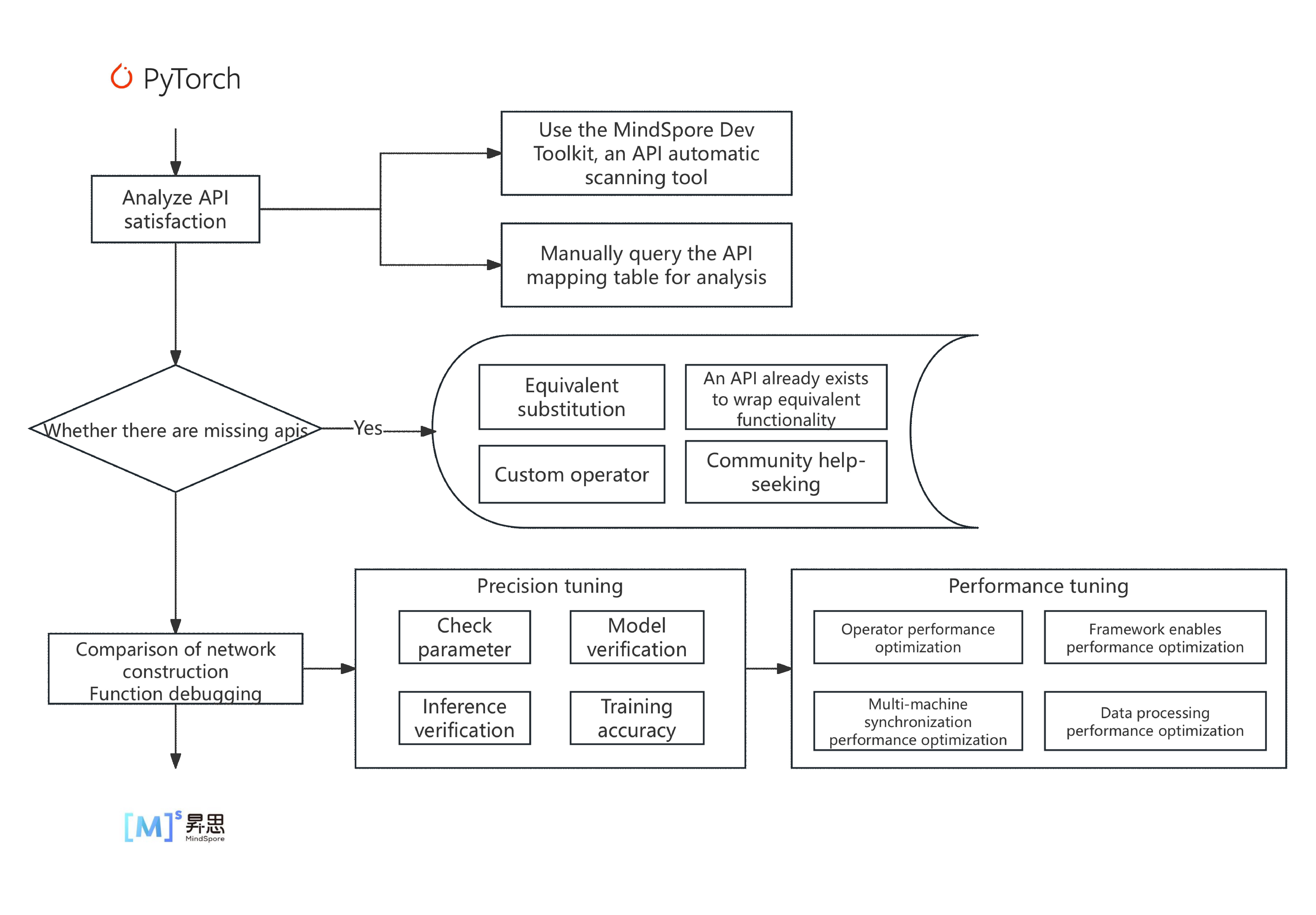}}
  \caption{The migration process from a model based on the Pytorch framework to the Mindspore framework}
\end{figure*}

\end{document}